\documentclass[
    ,final            % use final for the camera ready runs
    ,numberedheadings % uncomment this option for numbered sections
    ,cmfonts          % add further options here if necessary
  ]
  {aipproc}

\layoutstyle{6x9}

\begin{document}

\title{Cooling Evolution of Hybrid Stars}


\author{H.~Grigorian\footnote{Supported by the Virtual Institute of the
Helmholtz Association under grant No. VH-VI-041}\ }{
  address={Institut f\"ur Physik, Universit\"at Rostock,
D-18051 Rostock, Germany\\
E-mail: hovik.grigorian@uni-rostock.de},
 altaddress={Department of Physics,
Yerevan State University, 375025 Yerevan, Armenia}
}

\begin{abstract}
The cooling of compact isolated objects for different values of
the gravitational mass has been simulated for two alternative
assumptions. One is that the interior of the star is purely
hadronic\cite{bgv2004} and second that the star can have a rather
large quark core \cite{Grigorian:2004jq}. It has been shown that
within a nonlocal chiral quark model the critical density for a
phase transition to color superconducting quark matter under
neutron star conditions can be low enough for these phases to
occur in compact star configurations with masses below
$1.3~M_\odot$. For a realistic choice of parameters the equation
of state (EoS) allows for 2SC quark matter with a large quark gap
$\sim 100~$MeV for $u$ and $d$ quarks of two colors that coexists
with normal quark matter within a mixed phase in the hybrid star
interior.  We argue that, if in the hadronic phase the neutron
pairing gap in  $3P_2$ channel is larger than few keV and the
phases with unpaired quarks are allowed, the corresponding hybrid
stars would cool too fast.

Even in the case of the essentially suppressed $3P_2$ neutron gap
if free quarks occur for $M<1.3~M_\odot$, as it follows from our
EoS, one could not appropriately describe the neutron star cooling
data existing by today.

It is suggested to discuss a "2SC+X" phase, as a possibility to
have all quarks paired in two-flavor quark matter under neutron
star constraints, where the X-gap is of the order of 10 keV - 1
MeV. Density independent gaps do not allow to fit the cooling
data. Only the presence of an X-gap that decreases with increase
of the density could allow to appropriately  fit the data in a
similar compact star mass interval to that following from a purely
hadronic model.
\end{abstract}

\maketitle

%%%%%%%%%%%%%%%%%%%%%%%%%%%%%%%%%%%%%%%%%%%%
%% MAINMATTER
%%%%%%%%%%%%%%%%%%%%%%%%%%%%%%%%%%%%%%%%%%%%

\section{Introduction}

The ``standard'' scenario of neutron star
%%%(NS)
cooling is based on the main process responsible for the cooling,
which is the modified Urca process (MU) $nn\rightarrow
npe\bar{\nu}$ calculated using the free one pion exchange between
nucleons, see \cite{FM79}. However, this scenario explains only
the group of slow cooling data. To explain a group of rapid
cooling data ``standard'' scenario was supplemented by one of the
so called ``exotic'' processes either with pion condensate, or
with kaon condensate, or with hyperons, or involving the direct
Urca (DU) reactions, see \cite{T79,ST83} and refs therein. All
these processes may occur only for the density higher than a
critical density, $(2\div 6)~n_0$, depending on the model, where
$n_0$ is the nuclear saturation density. An other alternative to
''exotic'' processes is the DU process on quarks related to the
phase transition to quark matter.

Recently,  the cooling of neutron stars has been reinvestigated
within a purely hadronic model \cite{bgv2004}, i.e., when one
suppresses the possibility of quark cores in neutron star
interiors. We have demonstrated that the neutron star cooling data
available by today can be well explained within the {\em ''nuclear
medium cooling'' scenario}, cf. \cite{SVSWW97,V01}, i.e., if one
includes medium effects into consideration. In the ''standard plus
exotics'' scenario for hadronic models the in-medium effects have
not been incorporated, see \cite{TTTTT02,YGKLP03,PLPS04}. Recently
\cite{PLPS04} called this approach the ''minimal cooling''
paradigm. Some papers included an extra possibility of internal
heating that results in a slowing down of the cooling of old
pulsars, see \cite{T04} and Refs. therein.

The necessity to include in-medium effects into the neutron star
cooling is based on the whole experience of condensed matter
physics, see \cite{MSTV90,RW,IKHV01}. The relevance of in-medium
effects for the neutron star cooling problem has been shown by
\cite{V01,MSTV90,VS84,VS86,VS87} and the efficiency of the
developed ``nuclear medium cooling'' scenario for the description
of the neutron star cooling was demonstrated within the cooling
code by \cite{SVSWW97} and then by \cite{bgv2004}.

%% and takes into account a suppression of  the $3P_2$ neutron gap.
Each scenario puts some constraints on dense matter equation of
state (EoS).  In particular the density dependencies of the
asymmetry energy and the pairing gaps are the regulators of the
heat production and transport. The former dependence is an
important issue for the analysis of heavy ion collisions
especially within the new CBM (compressed baryon matter) program
to be realized at the future accelerator facility FAIR at GSI
Darmstadt.

The density dependence of the asymmetry energy also determines the
proton fraction in neutron star matter and thus governs the onset
of the very efficient direct Urca (DU) process. The DU process,
once occurring, would lead to a very fast cooling of neutron
stars. Within the ``standard + DU'' scenario the transition from
slow cooling to the rapid cooling occurs namely due to the
switching on the DU process. Thus the stars with $M< M_{\rm
crit}^{\rm DU}$ cool down slowly whereas the stars with the mass
only slightly above $M_{\rm crit}^{\rm DU}$ cool down very fast.
Since it is doubtful that many neutron stars belonging to an
intermediate cooling group have very similar masses, from our
point of view such a scenario seems unrealistic, cf.
\cite{bgv2004,KV04}. The modern EoS of the Urbana-Argonne group
\cite{APR98} allows for the DU process only for very high density
$n>5n_0$ (where $n_0$ is the saturation nuclear density) that
relates to the neutron star masses $M\geq M_{\rm crit}^{\rm
DU}\simeq 2 ~M_{\odot}$. Thus, using mentioned Urbana-Argonne
based EoS and the "standard +DU" scenario one should assume that
the majority of experimentally measured cooling points relates to
very massive neutron stars that seems us still more unrealistic.

The assumption about the mass distribution can be developed into a
more quantitative test of cooling scenarios when these are
combined with population synthesis models. The latter allow to
obtain Log N -- Log S distributions for nearby coolers which can
be tested with data from the ROSAT catalogue \cite{pgtb2004}.
Analysis \cite{pgtb2004} has supported ideas put forward in
\cite{bgv2004}.

At high star masses the central baryon density exceeds rather
large values $n>5n_0$. At these densities exotic states of matter
as, e.g., hyperonic matter or quark matter perhaps  are permitted.
Ref. \cite{Baldo:2003vx} argued that the presence of the quark
matter in massive compact star cores is a most reliable
hypothesis.

The possibility of the existence of neutron stars with large quark
matter cores is also not excluded
\cite{Grigorian:2004jq,bkv,ppls,bgv}.
%%that extend up to more than half of the star radius.
In the quark matter the DU process  yielding the rapid cooling may
arise on interacting but unpaired quarks \cite{Iwamoto:1980eb}

In this review we want to sketch a scenario for  the cooling of
hybrid stars.

\section{Structure of Hybrid Neutron Stars}

In describing the hadronic part of the hybrid star, as in
\cite{bgv2004}, we exploit a modification of the Urbana-Argonne
$V18+\delta v+UIX^*$ model of the EoS given in \cite{APR98}, which
is based on the most recent models for the nucleon-nucleon
interaction with the inclusion of a parameterized three-body force
and relativistic boost corrections. Actually we continue to adopt
an analytic parameterization of this model by Heiselberg and
Hjorth-Jensen \cite{HJ99}, hereafter HHJ.

The HHJ EoS  fits the symmetry energy  to the original Argonne
$V18+\delta v +UIX^*$ model in the mentioned density interval
yielding the threshold density for the DU process $n_c^{\rm
DU}\simeq~5.19~n_0$ ($M_c^{\rm DU}\simeq 1.839~M_{\odot}$).

We employ the EoS of a nonlocal chiral quark model developed in
\cite{BFGO} for the case of neutron star constraints with a
2-flavor color superconductivity (2SC) phase. It has been shown in
that work that the Gaussian formfactor ansatz leads to an early
onset of the deconfinement transition and such a model is
therefore suitable to discuss hybrid stars with large quark matter
cores \cite{GBA}.

The quark-quark interaction in the color anti-triplet channel is
attractive driving the pairing with a large zero-temperature
pairing gap $\Delta\sim 100$~MeV for the quark chemical potential
$\mu_q \sim (300\div 500)$~MeV, cf. \cite{arw98,r+98}, for a
review see \cite{RW} and references therein.
The attraction comes
either from the one-gluon exchange, or from a non-perturbative
4-point interaction motivated by instantons \cite{dfl}, or from
non-perturbative gluon propagators \cite{br}.

There may also exist a color-flavor locked (CFL) phase
\cite{arw99} for not too large values of the dynamical strange
quark mass or  large values of the baryon chemical potential
\cite{abr99}. In this phase the all quarks are paired. However,
the 2SC phase occurs at lower baryon densities than the CFL phase,
see \cite{SRP,NBO}. For applications to compact stars the omission
of the strange quark flavor is justified by the fact that chemical
potentials in central parts of the stars do barely reach the
threshold value at which the mass gap for strange quarks breaks
down and they appear in the system \cite{GBKG}.

 Following  Refs \cite{VYT02} we omit the possibility of
the hadron-quark mixed phase and found a tiny density jump on the
phase boundary from $n_c^{\rm hadr}\simeq 0.44~fm^{-3}$ to
{\bf{$n_c^{\rm quark}\simeq 0.46~fm^{-3}$.}}

\begin{figure}[ht]
%\vspace{-0.5cm}
  \includegraphics[height=0.6\textwidth,angle=-90]{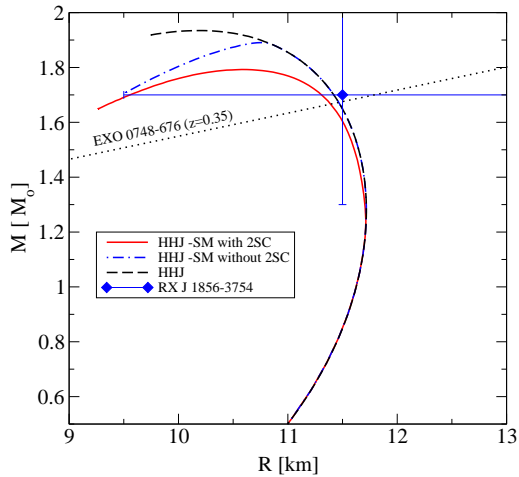}
   \caption{Mass -
radius relations for compact star configurations with different
EoS: purely hadronic star with HHJ EoS (dashed line), stable
hybrid stars with HHJ - Gaussian nonlocal chiral quark separable
model (SM) with 2SC phase (solid line) and with HHJ - SM, without
2SC phase (dash-dotted line). Data for two sources are also
indicated, see \cite{PLSP,CPM}. \label{fig:stab}}
\end{figure}

In Fig. \ref{fig:stab} we present the mass-radius relation for
hybrid stars with HHJ EoS vs. Gaussian nonlocal chiral quark
separable model (SM) EoS. Configurations for SM model, given by
the solid line, are stable, whereas without color super
conductivity (``HHJ-SM without 2SC'') no stable hybrid star
configuration is possible. In the case ``HHJ-SM with 2SC'' the
maximum neutron star mass proves to be $1.793~M_{\odot}$.

Additionally, within the ``HHJ-SM with 2SC'' phase we will allow
for the possibility of a weak pairing channel for all the quarks
which were unpaired, with typical gaps  $\Delta_X \sim 10$~keV
$\div 1$~MeV, as in the case of the CSL pairing channel, see
\cite{Schafer,Schmitt}. Since we don't know yet the exact pairing
pattern for this case, we call this hypothetical phase ``2SC+X''.
In such a way all the quarks get paired, some strongly in the 2SC
channel and some weakly in the X channel.

\section{Cooling}

We compute the neutron star thermal evolution adopting our fully
general relativistic evolutionary code. This code was originally
constructed for the description of hybrid stars by \cite{bgv}. The
main cooling regulators are the thermal conductivity, the heat
capacity and the emissivity. In order to better compare our
results with results of other groups we try to be as close as
possible to their inputs for the quantities which we did not
calculate ourselves. Then we add inevitable changes, improving
EoS.

%%%%%%%%%%%%%%%%%%%% Figure 4 %%%%%%%%%%%%%%%%%%
%\begin{figure}[htb]
%\includegraphics[height=0.85\textwidth,angle=-90]{TmTs}
%\caption{The relation between the inner crust temperature and the
%surface temperature for different models. Dash-dotted curves
%indicate boundaries of the uncertainty band. Notations of lines
%are determined in the legend. For more details see  \cite{bgv2004}
%and \cite{YLPGC03}. \label{T-in} }
%\end{figure}
%%%%%%%%%%%%%%%%%%%%%%%%%%%%%%%%%%%%%%%%%%%%%%%%

%\subsubsection{NS crust and Envelope}

The density $n\sim 0.5\div 0.7 ~n_0$ is the boundary of the
neutron star interior and the inner crust. The latter is
constructed of a pasta phase discussed by \cite{RPW83}, see also
recent works of \cite{MTVTCM04,T05}.

Further on we need the relation between the crust and the surface
temperature for neutron star. The sharp change of the temperature
occurs in the envelope.

\subsection{Cooling Evolution of Hadronic Stars}

Here we will shortly summarize the results on hadronic cooling.

In framework of ''minimal cooling'' scenario the pair breaking and
formation (PBF) processes may allow to cover an ''intermediate
cooling'' group of data (even if one artificially suppressed
medium effects)\cite{SVSWW97}. These processes are very efficient
for large pairing gaps, for temperatures being not much less than
the value of the gap.

%%%%%%%%%%%%%%%%%%%% Figure 5 %%%%%%%%%%%%%%%%%%
\begin{figure}[htb]
\includegraphics[height=0.6\textwidth,angle=-90]{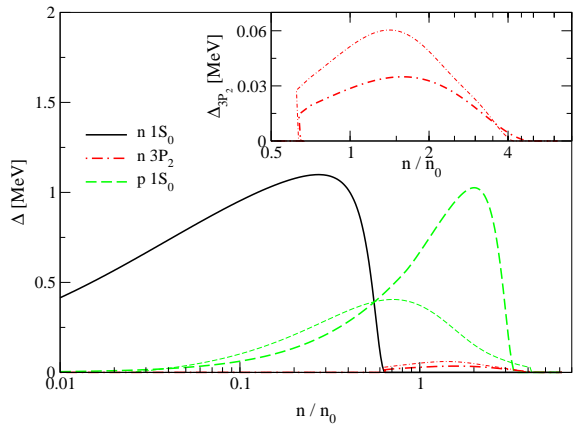} \caption{ Neutron
and proton pairing gaps according to model I
%%\cite{YGKLP03}
(thick solid, dashed and dotted lines) and according to model II
%%\cite{TT04}
%%%{SVSWW97}
(thin lines), see text. The $1S_0$ neutron gap is the same in both
models, taken from  \cite{AWP}. \label{fig-gaps}}
\end{figure}
%%%%%%%%%%%%%%%%%%%%%%%%%%%%%%%%%%%%%%%%%%%%%%%%

Gaps that we have adopted in the framework of the ''nuclear medium
cooling'' scenario, see \cite{bgv2004}, are presented in Fig.
\ref{fig-gaps}. Thick dashed lines show proton gaps which were
used in the work of \cite{YGKLP03} performed in the framework of
the ``standard plus exotics'' scenario. We will call the choice of
the ``3nt'' model from \cite{YGKLP03} the model I. Thin lines show
$1S_0$ proton and $3P_2$ neutron gaps from \cite{TT04}, for the
model AV18 by \cite{WSS95} (we call it the model II). Recently
\cite{SF03} has argued for a strong suppression of the $3P_2$
neutron gaps, down to values  $ \sim 10~$ keV, as the consequence
of the  medium-induced spin-orbit interaction.

These findings motivated \cite{bgv2004} to suppress values of
$3P_2$ gaps shown in Fig. \ref{fig-gaps} by an extra factor
$f(3P_2 ,n)=0.1$. Further possible suppression of the $3P_2$ gap
is almost not reflected on the behavior of the cooling curves.

Contrary to expectations of \cite{SF03} a more recent work of
\cite{KCTZ04} argued that the
%%$^3$P$_2$
$3P_2$ neutron pairing gap should be dramatically enhanced, as the
consequence of  the strong softening of the pion propagator.
According to their estimate, the
%%$^3$P$_2$
$3P_2$ neutron pairing gap is  as large as $1\div 10$~MeV in a
broad region of densities, see Fig. 1 of their work. Thus results
of calculations of \cite{SF03} and \cite{KCTZ04}, which both had
the same aim to include medium effects in the evaluation of the
$3P_2$ neutron gaps, are in a deep discrepancy with each other.

%\begin{figure}[ht]
%\vspace{-0.5cm}
%\includegraphics[height=0.9\textwidth,angle=-90]{EV_HJYmsn} \caption{Model
%I. Cooling curves according to the ''nuclear medium cooling''
%scenario, see Fig. 12 of \cite{bgv2004}. The labels correspond to
%the gravitational masses of the configurations in units of the
%solar mass.} \label{fig:cool-h}
%\end{figure}

%\begin{figure}[ht]
%\includegraphics[width=0.9\textwidth,angle=-90]{0404_f20}
%\caption{Model II. Same as
%%Fig. \ref{fig:cool-h} but
%using the Tsuruta law.
%%with the hadronic gaps taken from \cite{TT}, see
%Fig. 20 of Ref. \cite{bgv2004}.} \label{fig:cool-h-tt}
%\end{figure}

%%%%%%%%%%%%%%%%%%%%%%%%%%% Figure 15  %%%%%%%%%%%%%%%%%%
\begin{figure}[htb]
\includegraphics[height=0.65\textwidth,angle=-90]{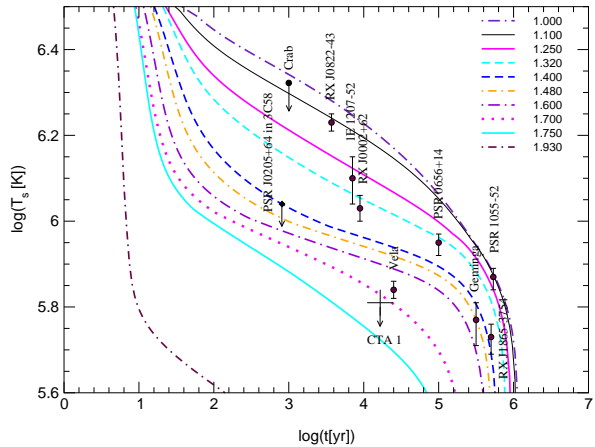}
%%EV_HJSms.ps
\caption{Fig. 21 of \cite{bgv2004}. Gaps are from Fig.
\ref{fig-gaps}
%%, thin line (Tamagaki and Takatsuka case).
for model II. The original $3P_2$ neutron
  pairing gap is  additionally suppressed by a factor $f(3P_2 ,n)=0.1$.
%%, as motivated by the result of  \cite{SF03}.
The $T_{\rm s} - T_{\rm in}$ relation is given by ``our fit''
curve of Fig. 4 in \cite{bgv2004}.
%Here and in all subsequent
%figures the value $T_{\rm s}$ is the red-shifted temperature. NS
%masses are indicated in the legend.
For more details see
\cite{bgv2004}. \label{fig21BGV}}
\end{figure}
%%%%%%%%%%%%%%%%%%%%%%%%%%%%%%%%%%%%%%%%%%%%%%%%
%\subsubsection{Summary of main conclusions for Hadronic cooling}
%Summarizing the main conclusions for Hadronic cooling we have:
\begin{itemize}

\item
Including superfluid gaps we see, in agreement with recent
microscopic findings of \cite{SF03},  that $3P_2$ neutron gap
should be as small as $ 10$~keV or less. So the ``nuclear medium
cooling'' scenario of \cite{bgv2004} supports results of
\cite{SF03} and
 fails to appropriately fit the neutron star
cooling data at the assumption of a strong enhancement of the
$3P_2$ neutron gaps as suggested by \cite{Grigorian:2005fi}.

%(we checked the range  $f(3P_2 ,n)=1\div 100$) and for moderately
%suppressed $1S_0$ proton gaps (for $f(1S_0 ,p)=0.1\div 0.5$). On
%the other hand the very same scenario allowed us to appropriately
%fit the whole set of data  at the assumption of a significantly
%suppressed $3P_2$ neutron gap (for $f(3P_2 ,n) \sim 0.1$). We
%observed an essential dependence of the results not only on the
%values of the gaps but also on their density dependence
%(see Fig.
%\ref{fig:cool-h})

\item
 Medium effects associated with the pion softening are called for
by the data. As the result of the pion softening the pion
condensation may occur for $n\geq n_c^{\rm PU}$ ($n\geq 3n_0$ in
our model). Its appearance at such rather high densities does not
contradict to the cooling data (see Fig. \ref{fig21BGV}), but also
the data are well described using the  pion softening but without
assumption on the pion condensation. This also means that the DU
threshold density can't be too low that puts restrictions on the
density dependence of the symmetry energy. Both  statements might
be important in the discussion of the heavy ion collision
experiments.

\item
 We demonstrated a regular mass dependence: for the neutron star masses $M>
1~M_{\odot}$ less massive neutron stars cool down slower, more
massive neutron stars cool faster.
\end{itemize}

\subsection{Cooling Evolution of Hybrid Stars with 2SC Quark Matter Core}

%%Below we also use another choice for the gaps (model II),  as shown by
%%dash lines in Fig. 5 of \cite{bgv2004}.
%%We will see that although the results are essentially model
%%dependent it does not change our qualitative conclusions.

For the calculation of the cooling of the quark core in the hybrid
star we use the model \cite{bgv}. We incorporate the most
efficient processes: the quark direct Urca (QDU) processes on
unpaired quarks, the quark modified Urca (QMU), the quark
bremsstrahlung (QB), the electron bremsstrahlung (EB), and the
massive gluon-photon decay (see \cite{bkv}). Following \cite{JP02}
we include the emissivity of the quark pair formation and breaking
(QPFB) processes. The specific heat incorporates the quark
contribution, the electron contribution and the massless and
massive gluon-photon contributions. The heat conductivity contains
quark, electron and gluon terms.

The calculations are based on the hadronic cooling scenario
presented in Fig. \ref{fig21BGV} and we add the contribution of
the quark core. For the Gaussian form-factor the quark core occurs
already for $M>1.214~M_\odot$ according to the model \cite{BFGO},
see Fig. \ref{fig:stab}. Most of the relevant neutron star
configurations (see Fig. \ref{fig21BGV}) are then affected by the
presence of the quark core.

\begin{figure}
%\vspace{-0.5cm}
\includegraphics[height=0.65\textwidth,angle=-90]{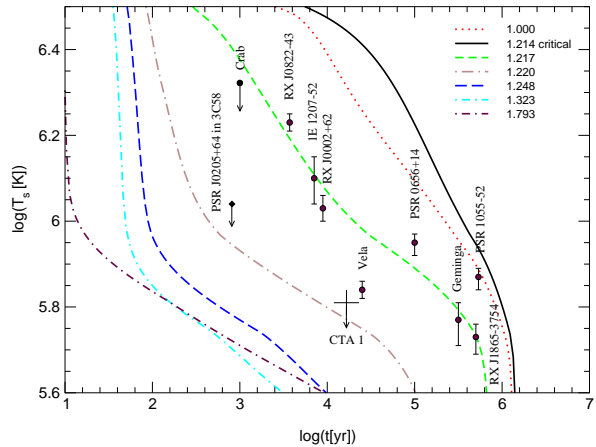}
\caption{Model I. Cooling curves for hybrid star configurations
with Gaussian quark matter core in the 2SC phase. The labels
correspond to the gravitational masses of the configurations in
units of the solar mass.} \label{fig:cool-2sc}
\end{figure}

First we check the possibility of the 2SC+ normal quark phases
Fig. \ref{fig:cool-2sc}.

The variation of the gaps for the strong pairing of quarks within
the 2SC phase and the gluon-photon mass in the interval $\Delta,
m_{g-\gamma}\sim 20\div 200~$MeV only slightly affects the
results. The main cooling process is the QDU process on normal
quarks. We see that the presence of normal quarks entails too fast
cooling. The data could be explained only if all the masses lie in
a very narrow interval ($1.21<M/M_\odot<1.22$ in our case). In
case of the other two crust models the resulting picture is
similar.

%The value $M\simeq ???$ depends on the model for the form-factor
%and can be shifted. However
The existence of only a very narrow mass interval in which the
data can be fitted seems us unrealistic as by itself as from the
point of view of the observation of the neutron stars in binary
systems with different masses, e.g., $M_{\rm B1913+16}\simeq
1.4408 \pm 0.0003~M_{\odot}$ and $M_{\rm J0737-3039B}\simeq 1.250
\pm 0.005~M_{\odot}$, cf. \cite{L04}. {\em{Thus the data can't be
satisfactorily explained.}}

We first check the case  $\Delta_X$ to be constant. For the
$\Delta_X \simeq 1~$MeV  cooing  is too slow
\cite{Grigorian:2004jq}. It is true for all three crust models.
Thus the gaps for formerly unpaired quarks should be still smaller
in order to obtain a satisfactory description of the cooling data.

For the $\Delta_X = 30~$ keV the cooling data can be  fitted but
have a very fragile dependence on the gravitational mass of the
configuration. Namely, we see that all data points, except the
Vela, CTA 1 and Geminga, correspond to hybrid stars with masses in
the narrow interval $M=1.21\div 1.22 ~M_\odot$

Therefore we would like to explore whether a density-dependent
X-gap could allow a description of the cooling data within a
larger interval of compact star masses.

\begin{figure}[ht]
\includegraphics[height=0.65\textwidth,angle=-90]{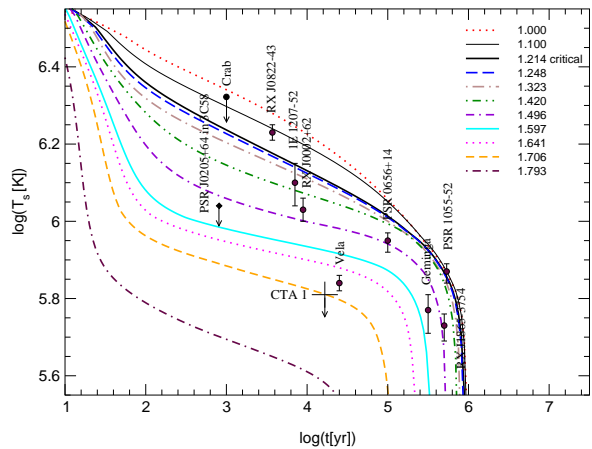}
\caption{ Cooling curves for hybrid star configurations with
Gaussian quark matter core in the 2SC phase with a density
dependent pairing gap according to Eq. (\ref{gap}) for model II. }
\label{fig:cool-csl-x-tt}
\end{figure}

We employ the ansatz: X-gap as a decreasing function of the
chemical potential
\begin{equation}
\label{gap}
\Delta_X(\mu)=\Delta_c~\exp[-\alpha(\mu-\mu_c)/\mu_c]~,
\end{equation}
where the parameters are chosen such that at the critical quark
chemical potential $\mu_c=330$ MeV for the onset of the
deconfinement phase transition the X-gap has its maximal value of
$\Delta_c=1.0$ MeV and at the highest attainable chemical
potential $\mu_{\rm max}=507$ MeV, i.e. in the center of the
maximum mass hybrid star configuration it falls to a value of the
order of $10$ keV. We choose the value $\alpha=10$ for which
$\Delta_X(\mu_{\rm max})=4.6$ keV. In Fig. \ref{fig:cool-csl-x-tt}
we show the resulting cooling curves for the gap model II with gap
anzatz \ref{gap}, which we consider as the most realistic one.

We observe that the mass interval for compact stars which obey the
cooling data constraint ranges now from $M=1.32~M_\odot$ for slow
coolers up to $M=1.75~M_\odot$ for fast coolers such as Vela, cf.
with that we have found with the purely hadronic model
\cite{bgv2004} with different parameter choices. Note that
according to a recently suggested independent test of cooling
models \cite{pgtb2004} by comparing results of a corresponding
population synthesis model with the Log N - Log S distribution of
nearby isolated X-ray sources the cooling model I did not pass the
test. Thereby it would be interesting to see whether our quark
model within the gap ansatz II
%%%(\ref{gap}) may
could pass the  Log N - Log S test.

%\begin{figure}[ht]
%\vspace{-0.5cm}
%\includegraphics[height=0.9\textwidth,angle=-90]{EV_HJY_2sc_xCn}
%\caption{Model I. Cooling curves for hybrid star configurations
%with Gaussian quark matter core in the 2SC phase with a density
%dependent pairing gap according to Eq. (\ref{gap}).}
%\label{fig:cool-csl-x}
%\end{figure}

\section{Conclusion}

\begin{itemize}
\item
Within a nonlocal, chiral quark model the critical densities for a
phase transition to color superconducting quark matter  can be low
enough for these phases to occur in compact star configurations
with masses below $1.3~M_\odot$.

\item
For the choice of the Gaussian form-factor the 2SC quark matter
phase arises at $M\simeq 1.21~M_\odot$.

\item
Without a residual pairing the 2SC quark matter phase could
describe the cooling data only if compact stars had masses in a
very narrow band around the critical mass for which the quark core
can occur.

\item

Under assumption that formally unpaired quarks can be paired with
small gaps $\Delta_X <1~$MeV  (2SC+X pairing), which values we
varied in wide limits, only for density dependent gaps the cooling
data can be appropriately fitted.
\end{itemize}

So the present day cooling data could be still explained by hybrid
stars, however, when assuming a complex pairing pattern, where
quarks are partly strongly paired within the 2SC channel, and
partly weakly paired with gaps $\Delta_X < 1~$MeV, being rapidly
decreasing with the increase of the density.

It remains to be investigated which microscopic pairing pattern
could fulfill the constraints obtained in this work. Another
indirect check of the model could be the Log N - Log S test.

\begin{theacknowledgments}
The research has been supported by the Virtual Institute of the
Helmholtz Association under grant No. VH-VI-041 and by the DAAD
partnership programme between the Universities of Yerevan and
Rostock. In particular I acknowledge D. Blaschke for his active
collaboration and support. The results reported in these
Proceedings are obtained in collaboration with my colleagues D.
Blaschke, D.N. Voskresensky and D.N. Aguilera.  I thank the
organizers of Spa HLPR2004 meeting.
\end{theacknowledgments}

\end{document}